\documentstyle[aps,eqsecnum]{revtex}

\def\lb#1{\label{#1}}
\def\l#1{\lb{#1}}
\def\r#1{(\ref{#1})}
\def\c#1{\cite{#1}}

\def\beq{\begin{equation}}
\def\eeq{\end{equation}}
\def\bez{\begin{displaymath}}
\def\eez{\end{displaymath}}
\def\beb#1\l#2\eeb{\begin{equation}
\begin{array}{c} #1 
\end{array} \label#2  \end{equation}}
\def\bey#1\eey{\begin{displaymath}
\begin{array}{c} #1  \end{array}  \end{displaymath}}

\begin{document}

\begin{titlepage}
\begin{flushright}
November 29, 2001\\
hep-th/0111270
\end{flushright}

\begin{centering}
\vfill
{\bf 
ON CORRESPONDENCE   OF   BRST-BFV,   DIRAC   AND   REFINED   ALGEBRAIC
QUANTIZATIONS OF CONSTRAINED SYSTEMS
}
\vspace{1cm}

O. Yu. Shvedov \footnote{shvedov@qs.phys.msu.su} \\
\vspace{0.3cm}
{\small {\em Sub-Dept. of Quantum Statistics and Field Theory,}}\\
{\small{\em Department of Physics, Moscow State University }}\\
{\small{\em Vorobievy gory, Moscow 119899, Russia}}

\vspace{0.7cm}

\end{centering}

{\bf Abstract}

Correspondence between BRST-BFV,  Dirac and refined  algebraic  (group
averaging,  projection  operator)  approaches  to quantize constrained
systems is analyzed.  For the closed-algebra case,  it is  shown  that
the  component  of  the BFV wave function with maximal (minimal)
number of ghosts and antighosts in the Schrodinger representation  may
be  viewed  as  a  wave  function  in  the  refined  algebraic (Dirac)
quantization approach.  The Giulini-Marolf group averaging formula for
the  inner  product  in the refined algebraic quantization approach is
obtained from the  Batalin-Marnelius  prescription  for  the  BRST-BFV
inner  product  which  should be generally modified due to topological
problems. The considered prescription for the correspondence of states
is  observed  to  be  applicable  to  the  open-algebra case.  Refined
algebraic quantization approach is generalized then  to  the  case  of
nontrivial   structure  functions.  A  simple  example  is  discussed.
Correspondence of observables in  different  quantization  methods  is
also investigated.

\vspace{0.7cm}

\vfill \vfill
\noindent

\end{titlepage}
\newpage
\sloppy

\section{Introduction}

Theory  of constrained systems is a basis of modern physics:
gauge field theories,  quantum gravity and  supergravity,  string  and
superstring models are examples of systems with constraints.  For such
theories, one should  specify  not  only  an  evolution  equation  but
additional requirements  (constraints) imposing on initial conditions.
Alternatively, one can take the constraints into account by  modifying
the inner product of the theory.  For some cases, Hamiltonian is zero,
so  that  all  the  dynamics  is  involved  to  constraints,  and  the
well-known  problem  of  time  in reparametrization-invariant theories
arises.

Different approaches have been developed  for  quantizing  constrained
systems. Each   procedure   of   quantization   has   advantages   and
disadvantages. The most difficult problem is  to  introduce  an  inner
product.

Quantization approaches  can  be  divided  into  two classes:  without
introducing additional degrees of freedom and with introducing ghosts,
antighosts, Lagrange  multipliers  and canonically conjugated momenta.
The latter  class  of  quantization  methods  is  known  as   BRST-BFV
approach \c{BRST,BFV,KO} (for a review see \c{Henneaux}).

Examples of  quantization  techniques of  the  first  class  are Dirac
method \c{D} (including gauge-fixing  approach  \c{D,Faddeev}) and  refined
algebraic quantization    (group   averaging   method)   \c{Marolf,M2}
(analogous ideas   are   used   in   projection   operator    approach
\c{proj,Shabanov} and  in  lattice  gauge  theories \c{lattice}).  The
Dirac quantization is based on imposing the constraint  conditions  on
physical  states.  It is not easy to introduce an inner product in the
Dirac approach.  However,  the refined algebraic  quantization  method
allows us to overcome this difficulty: instead of imposing constraints
on physical states,  one modifies the physical inner  product  due  to
constraints.

By now,  refined algebraic quantization has  been  developed  for  the
closed algebras   of   constraints  only.  It  has  been  stressed  in
\c{Marolf3} that  generalization  of  this  method  to  the  case   of
nontrivial structure functions is an interesting open problem.

The purposes of this paper are:

- to  generalize  the  refined  algebraic quantization approach to the
case of constraint algebras with nontrivial structure functions;

- to investigate the correspondence between states and observables  in
Dirac, refined algebraic and BRST-BFV quantization approaches.

This paper  is organized as follows.  In section 2 the Dirac,  refined
algebraic and BRST-BFV quantization approaches are  reviewed.  General
requirements for   the   refined   algebraic  inner  product  for  the
open-algebra case are formulated.  Section 3 deals  with  finding  the
correspondence between states in BRST-BFV, Dirac and refined algebraic
quantization approaches.  The  obtained  formulas  are  justified  for
different known  prescriptions  of  the BRST-BFV inner product for the
closed-algebra case.  This correspondence is supposed to be valid  for
the open-algebra case as well.  This allows us to obtain a formula for
the refined algebraic quantization inner product for the  open-algebra
case in section 4. The correspondence between observables in different
quantization approaches is discussed in section  5.  Section  6  deals
with  a  simple  example  of  a  constrained  system  with  nontrivial
structure functions. Section 7 contains concluding remarks.

\section{Quantization methods}

\subsection{Dirac approach}

The most famous way to quantize a  constrained  system  is  the  Dirac
approach \c{D}. It is as follows.
Evolution transformation is
presented as $\exp(-iH^+t)$ for some Hamiltonian $H$.  Usually, $H$ is
Hermitian, $H^+  =  H$.  However,  the   examples   of   non-Hermitian
Hamiltonians are  presented below.  Constraints are taken into account
by imposing the additional conditions:
\beq
\hat{\Lambda}_a^+ \Psi = 0, \qquad a=\overline{1,M},
\l{1}
\eeq
on the states. The operators
$\hat{\Lambda}_a^+$   are   quantum   analogs   of  constraints.
For the closed-algebra unimodular case, constraints are Hermitian,
$\hat{\Lambda}_a^+ = \hat{\Lambda}_a$.
Requirements \r{1} do not contradict each  other,  provided  that  the
constraints commute  on the constraint surface.  For the quantum case,
this means that
\beq
[\hat{\Lambda}_a^+, \hat{\Lambda}_b^+]      =      i      (U^c_{ab})^+
\hat{\Lambda}_c^+
\l{2}
\eeq
for some operators $U^c_{ab}$ called usually  as  structure  functions
\c{Henneaux}. Relation   \r{1}   should   also   conserve  under  time
evolution, so that constraints should commute with the Hamiltonian for
the states obeying eq.\r{1}:
\beq
[H^+, \hat{\Lambda}_a^+] = i (R^c_a)^+ \hat{\Lambda}_c^+
\l{3}
\eeq
for some operators $R^c_a$.

It is not easy to construct an inner product  in  the  Dirac  approach
since $\Psi(q)$   are   distributions  rather  than  square-integrable
functions.

{\it Example  1.} Consider the case $M=1$,  $q=x$,  $\hat{\Lambda} = -
i\frac{\partial}{\partial x}$. Then condition \r{1} will take the form
$\frac{\partial \Psi}{\partial x} = 0$,  so that $\Psi = const$ is not
square integrable.

One   usually   imposes   additional    gauge    conditions
\c{D,Henneaux,Faddeev} in  such  a way that each gauge orbit should be
taken into  account  once (for example 1,  such gauge condition may be
chosen as $x=0$).  The wave function  $\Psi(q)$  is  considered
then on  the  gauge  surface only,  since the off-gauge-surface values
$\Psi$ are specified from the constraint  equations  \r{1};  then  the
integral of   $|\Psi|^2$   is  taken  over  the  gauge  surface  only.
Unfortunately, this approach is gauge-dependent,  especially  for  the
case of the Gribov copies problem \c{Gribov,Shabanov}.

\subsection{Refined algebraic quantization approach}

An alternative way to develop the quantum theory is to use the refined
algebraic approach \c{Marolf,M2} and take the constraints into account
by modifying the inner product instead of imposing requirements \r{1}.
States are specified by  smooth  and  damping  at  the  infinity  wave
functions $\Phi(q)$  called  auxiliary state vectors.  However,  their
inner product is given  by  a  nontrivial  formula.  Let  us  consider
abelian and nonabelian cases.

\subsubsection{Abelian case}

For the  simplest  abelian  case ($\hat{\Lambda}_a^+=\hat{\Lambda}_a$,
$U^c_{ab} = 0$) and constraints  with  continuous  spectra,  the  inner
product reads:
\beq
(\Phi, \prod_a 2\pi \delta(\hat{\Lambda}_a) \Phi).
\l{3-1}
\eeq
Since it is degenerate,  one should factorize the space  of  auxiliary
states ${\cal H}_{aux}$:  wave functions $\Phi_1$ and $\Phi_2$ are set
to be equivalent if $(\Phi,  \prod_a 2\pi\delta(\Lambda_a)  (\Phi_1  -
\Phi_2)) =  0$  for  all $\Phi$.  The corresponding factorspace
${\cal H}_{aux}/\sim$ should be completed to obtain a physical Hilbert
state space $\overline{{\cal H}_{aux}/\sim}$.

In particular, the following transformation
\beq
\Phi \to \Phi + \hat{\Lambda}_a X^a
\l{4-1}
\eeq
takes a auxiliary state to the equivalent state and can be called as a
quantum gauge transformation.

For example 1, formula \r{3-1} for the inner product takes the form
$|\int_{-\infty}^{+\infty} dx \Phi(x)|^2$.
Two auxiliary  functions  are then equivalent if their integrals $\int
dx \Phi(x)$ coincide.  The classes of  equivalence  are  specified  by
numbers $\int  dx  \Phi(x)$,  so  that  the  physical Hilbert space is
trivial.

Formulas of  the  Dirac  approach are indeed reproduced in the refined
algebraic quantization approach.  For each auxiliary  state,  consider
the wave function (distribution) \c{Marolf}
\beq
\Psi = \prod_a 2\pi\delta(\hat{\Lambda}_a) \Phi.
\l{5-1}
\eeq
For equivalent auxiliary states,  we obtain  the  same  $\Psi$.  Thus,
physical states  can  be specified by distributions $\Psi$ obeying the
Dirac condition \r{1}. Moreover, we see that the inner product for the
Dirac states  is  introduced.  For  Dirac  wave  functions $\Psi_1$ and
$\Psi_2$ satisfying \r{1},  one should find $\Phi_1$ and $\Phi_2$ from
relation \r{5-1}   and   evaluate   the   quantity  $(\Phi_1,  \prod_a
2\pi\delta(\hat{\Lambda}_a) \Phi_2)$.  This result will not depend  on
the particular  choice of representatives $\Psi_1$ and $\Psi_2$ of the
equivalence classes.

For example  1,  formula  \r{5-1} is taken to the  form $\Psi(x) = \int dy
\Phi(y)$. We  see  that  $\Psi$  is  $x$-independent.  Moreover,   for
constant functions  $\Psi_1$  and $\Psi_2$ we find their inner product
$\Psi_1^*\Psi_2$.

\subsubsection{Nonabelian case}

The nonabelian case is more  complicated  for  the  refined  algebraic
quantization approach  \c{M2}.  The  Hermitian  parts  of  constraints
$\check{\Lambda}_a$ ($\check{\Lambda}_a^+    =     \check{\Lambda}_a$)
satisfy the following closed-algebra relations:
\beq
[\check{\Lambda}_a; \check{\Lambda}_b] = i f^c_{ab} \check{\Lambda}_c.
\l{1-1}
\eeq
for some structure constants $f^c_{ab}$. Let
$L_a$, $a=\overline{1,M}$ be generators of the Lie  algebra  with  the
following commutation relations $[L_a,L_b]=if_{ab}^cL_c$. Consider the
corresponding  Lie  group  $G$  and  the exponential mapping $\mu^aL_a
\mapsto \exp(i\mu^aL_a)$.  The operators  $\check{\Lambda}_a$  form  a
representation of      the      Lie       algebra,       so       that
$\exp(i\mu^a\check{\Lambda}_a)$  will  form  a representation of group
$\check{T}(\exp(i\mu^aL_a)) = \exp(i\mu^a\check{\Lambda}_a)$.  By  $Ad
(L_a)$  we denote the adjoint representation of the Lie algebra,  $(Ad
(L_a) \rho)^c = if^c_{ab}  \rho^b$,  while  $Ad\{g\}$  is  an  adjoint
representation  of the group $(Ad\{g\} \rho)^c = (\exp(A))^c_b \rho^b$
with $A^c_b = - \mu^a f^c_{ab}$, $g = \exp(i\mu^aL_a)$.

For the general closed-algebra case,  the inner product  is  expressed
via  the integral over gauge group with the help of the Giulini-Marolf
group averaging formula \c{M2} instead of \r{3-1}:
\beq
\int d_Rg (det Ad \{g\})^{-1/2} (\Phi, \check{T}(g) \Phi)
\l{16-1}
\eeq
Here $d_Rg$ is the right-invariant Haar measure on the group $G$.

Formula \r{3-1} is indeed a partial case of \r{16-1}.  For the abelian
case considered above, one has $d_Rg = d\mu^1...d\mu^M$, $\check{T}(g)
= \exp\{i\mu^a\check{\Lambda}_a\}  =   \exp\{i\mu^a\hat{\Lambda}_a\}$,
$\det Ad\{g\} = 1$, so that formula \r{16-1} takes the form
\beq
\int
d\mu^1...d\mu^M
(\Phi, \exp\{i\mu^a\hat{\Lambda}_a\} \Phi)
\l{1-2}
\eeq
Integrating over $\mu$, we obtain formula \r{3-1}.

The case of discrete spectrum of constraints can  be  also  considered
within framework of eq.\r{16-1}.

{\it Example 2.} Let $M=1$, $q=\varphi \in (0,2\pi)$, $\hat{\Lambda} =
- i \frac{\partial}{\partial\varphi}$,  the wave  functions  obey  the
periodic  boundary  conditions $\Phi(\varphi + 2\pi) = \Phi(\varphi)$.
Formula \r{1-2}  takes  the  form  $\int_0^{2\pi}  d\varphi  \int d\mu
\Phi^*(\varphi) \Phi(\varphi + \mu)$. If one performed the integration
over $|mu  \in  (-\infty,  +  \infty)$,  the  inner  product  would be
divergent. However, one should take into account that
$\exp\{2\pi i\hat{\Lambda}\} = 1$. Therefore,  the gauge group  is  $U(1)$,
so that   the  integration in eq.\r{1-2} should be performed only over $\mu
\in (0,2\pi)$.  For the inner product \r{1-2},  we obtain then formula
$|\int_0^{2\pi}  d\varphi  \Phi(\varphi)|^2$  which  is a basis of the
projection operator quantization \c{proj}.

Two  states  $\Phi_1$  and  $\Phi_2$  are called
gauge-equivalent if their difference satisfies the condition
\bez
\int d_Rg (det Ad \{g\})^{-1/2} \check{T}(g) (\Phi_1-\Phi_2) = 0.
\eez
For example, states
$X$ and $(det Ad \{h\})^{-1/2} \check{T}(h) X$  are  equivalent.  This
means that
\bez
[(det Ad \{h\})^{-1/2} \check{T}(h) - 1] X  \sim 0.
\eez
After substitution $h = \exp(i\rho^aL_a)$ we find in the leading order
in $\rho$ that
\beq
\hat{\Lambda}_a X \sim 0
\l{17-1}
\eeq
with
\beq
\hat{\Lambda}_a = \check{\Lambda}_a - \frac{i}{2} f^b_{ab}
\l{2-2}
\eeq
The fact that constraints in the non-unimodular  case  should  be  not
Hermitian was discussed in \c{M2,KS} in details.

The Dirac wave function can be specified as
\beq
\Psi = \int d_Rg (det Ad \{g\})^{-1/2} T(g)\Phi.
\l{17a-1}
\eeq
analogously to eq.\r{5-1}.
It obeys the condition
\c{M2}
\bez
(det Ad \{h\})^{1/2} \check{T}(h)\Psi = \Psi
\eez
which can be also presented in the infinitesimal form
\beq
\tilde{\Lambda}_a^+ \Psi \equiv  (\Lambda_a  +  \frac{i}{2}  f_{ab}^b)
\Psi = 0
\l{18-1}
\eeq
found in \c{KS}.

\subsubsection{General requirements   for   the  nontrivial  structure
functions case}

We are going to generalize the refined algebraic quantization approach
to  the  case  of  nontrivial  structure  functions  $U^c_{ab}$  being
operators rather than constants.  It has been stressed in  \c{Marolf3}
that  generalization  of  the  Giulini-Marolf  formula \r{16-1} to the
open-algebra case is an interesting open problem.

One can hope that the inner product formula for the  auxiliary  states
should be looked for in the following form
\beq
(\Phi, \eta \Phi)
\l{4}
\eeq
for some operator $\eta$ such that $\eta^+ = \eta$,  $\eta \ge 0$. The
main requirement for the operator $\eta$ is
\beq
\eta \hat{\Lambda}_a = 0.
\l{5}
\eeq
Two auxiliary states are called equivalent if their difference $\Delta
\Phi$ has zero norm, this is certainly the case if
\bez
\Delta \Phi = \hat{\Lambda}_a Y^a
\eez
The classes  of  equivalence  being elements of the factorspace
${\cal H}_{aux}/\sim$
correspond to the Dirac states with the help of formula
\beq
\Psi = \eta \Phi.
\l{6}
\eeq
The constrained conditions \r{1} are automatically satisfied then. The
space of  physical  states  is  defined  as  a  completeness  of   the
factorspace $\overline{{\cal H}_{aux}/\sim}$.

However, such a generalization is not trivial.  To find it, it will be
necessary to
discuss a   relationship  between  the  Dirac  and  refined  algebraic
quantization approaches and BRST-BFV quantization technique.

\subsection{BRST-BFV approach}

To develop  the  BRST-BFV  approach  \c{BRST,BFV,KO,Henneaux},  it  is
necessary to introduce additional degrees of freedom:
Lagrange multipliers  and  momenta
$\lambda^a$, $\pi_a$,  $a=\overline{1,M}$, ghosts and antighosts
$C^a, \overline{C}_a$, canonically conjugated momenta
$\overline{\Pi}_a, \Pi^a$,    $a=\overline{1,M}$.    The    nontrivial
(anti)commutation relations are:
\bez
[\lambda^a, \pi_b] = i\delta^a_b,  \qquad  [C^a,  \overline{\Pi}_b]_+  =
\delta^a_b, \qquad [\overline{C}_a, \Pi^b]_+ = \delta_a^b
\eez
Operators $\overline{C}_a$ and $\Pi^b$ are anti-Hermitian,  others are
Hermitian. The  main  object  of the BRST-BFV method is the $B$-charge
$\Omega$. For the closed-algebra case, it has the form
\beq
\Omega = C^a\check{\Lambda}_a
- \frac{i}{2} f^a_{bc} \overline{\Pi}_a C^bC^c -
\frac{i}{2} f^a_{ba} C^b - i\pi_a\Pi^a.
\l{6-1}
\eeq
It is formally Hermitian and nilpotent,
\beq
\Omega^+ = \Omega; \qquad \Omega^2 = 0.
\l{3-2}
\eeq
For the  open-algebra  case  with nontrivial structure functions,  the
B-charge is looked for in the following form:
\beq
\Omega =   -   i\pi_a   \Pi^a   +   C^a   \hat{\Lambda}_a   +   ...  +
\Omega^n{}^{b_1...b_{n-1}}_{a_1...a_n} \overline{\Pi}_{b_1}        ...
\overline{\Pi}_{b_{n-1}} C^{a_1} ... C^{a_n} + ...
\l{7}
\eeq
The operators $\overline{\Pi}$ and $C$ are ordered in formula \r{7} in
such a way that ghosts $C$ are put to the right, while the momenta
$\overline{\Pi}$ are put to the left.
The operator-valued coefficient functions
$\Omega^n{}^{b_1...b_{n-1}}_{a_1...a_n}$
being antisymmetric separately with respect to  $b_1,...,b_{n-1}$  and
separately with respect to $a_1,...,a_n$ are constructed in a standard
way \c{Henneaux} from recursive relations that are corollaries of  the
properties \r{3-2}.
Formula \r{7}  is  in  agreement  with  relation  \r{2-2}  since   the
coefficient of $C^a$ in eq.\r{6-1} is indeed $\hat{\Lambda}_a$.

Instead of  requirement  \r{1},  the  BRST-BFV condition is imposed on
physical states $\Upsilon$:
\beq
\Omega\Upsilon = 0,
\l{9}
\eeq
The gauge freedom is also allowed, the gauge transformation is
\beq
\Upsilon \to \Upsilon + \Omega X,
\l{10}
\eeq
so that  states  $\Upsilon$ and $e^{[\Omega,\rho]_+}\Upsilon$ are also
equivalent.

Another requirement  is  that  physical states should be of zero ghost
number,
\bez
N = \Pi^a \overline{C}_a - \overline{\Pi}_a C^a,
\eez
so that
\beq
N\Upsilon = 0.
\l{4-2}
\eeq

The most nontrivial problem is to introduce an inner  product  in  the
BRST-BFV formalism.
Consider  the Schrodinger
representation for the BFV wave function $\Upsilon$,
$\Upsilon = \Upsilon (q,\lambda,\Pi,\overline{\Pi})$.
The operators are rewritten then as
\beq
C^a = \frac{\partial}{\partial \overline{\Pi}_a};
\qquad
\overline{C}_a = \frac{\partial}{\partial {\Pi}^a};
\qquad
\pi_a = - i\frac{\partial}{\partial\lambda^a};
\qquad
p_i = - i\frac{\partial}{\partial q^i},
\l{14a}
\eeq
the left derivatives are considered here.
The inner product is indefinite. Formally, it is as follows  \c{JMP0}
\beq
(\Upsilon_1,\Upsilon_2) =     \int     dq     \prod_{a=1}^M     d\mu^a
d\overline{\Pi}_a d\Pi^a (\Upsilon_1(q,i\mu,\Pi,\overline{\Pi}))^*
\Upsilon_2(q,-i\mu,\Pi,\overline{\Pi}).
\l{12-1}
\eeq
The integration and conjugation rules are 
$(\overline{\Pi}_{a_1}...\overline{\Pi}_{a_l} \Pi^{b_1} ... \Pi^{b_s})^* = 
(-1)^s \Pi^{b_s}... \Pi^{b_1} \overline{\Pi}_{a_l} ... \overline{\Pi}_{a_1}$, 
$\int d\overline{\Pi}_a \overline{\Pi}_a =1$, 
$\int d{\Pi}^a {\Pi}^a =1$.  However, the inner product space \r{12-1}
requires additional investigation. For example, a class of allowed BFV
wave functions $\Upsilon$ should be specified.

For the abelian case, one introduces the
creation and annihilation operators
\beq
A_a^{\pm} = \frac{1}{\sqrt{2}} [\pi_a \pm iM_a{}^b \Lambda_b]
\l{5-2}
\eeq
for some Hermitian real positively definite nondegenerate matrix  $M$,
shows \c{Razumov}
that it is possible to perform such a gauge transformation \r{10}
that after it
\beq
A_a^- \Upsilon = 0.
\l{14-1}
\eeq
Then the inner product \r{12-1} is shown to be convergent \c{Razumov}.

The more  general  formula  for  the  inner  product  was  written  in
\c{Marnelius}. First,   one   considers  the  representatives  of  the
equivalence classes which obey the following additional conditions
\beq
C^a\Upsilon = 0, \qquad \pi_a \Upsilon = 0
\l{9-1}
\eeq
which make  the  state $\Upsilon$ BRST-BFV-invariant.  Unfortunately,
the quantity  $(\Upsilon,\Upsilon)$  is  ill-defined.   However,   the
expression
\beq
(\Upsilon, e^{t[\Omega,\rho]_+} \Upsilon)
\l{10-1}
\eeq
which is formally equivalent to  $(\Upsilon,\Upsilon)$  occurs  to  be
well-defined for a certain choice of the gauge fermion $\rho$,
\beq
\rho = - \lambda^a \overline{\Pi}_a.
\l{10aa-1}
\eeq

Let us  analyze  the prescriptions for the inner products of different
quantization methods  and  find  a   correspondence   between   states
$\Upsilon$, $\Phi$, $\Psi$.

\section{Correspondence of states}

\subsection{Abelian case}

Let us   investigate   the  inner  product  \r{12-1}  under  condition
\r{14-1}. Relation \r{5-2} and $B$-condition \r{9} imply that
\bez
[\frac{\partial}{\partial \overline{\Pi}_a}  + M_b{}^a \Pi^b] \Upsilon
=0
\eez
so that
\bez
\Upsilon(q,\lambda,\Pi,\overline{\Pi}) =    \exp    [-\overline{\Pi}_a
M_b{}^a \Pi^b] \Upsilon_0(q,\lambda).
\eez
Condition \r{14-1} implies that
\bez
\Upsilon_0(q,\lambda) = \exp [\lambda^a M_a{}^b \Lambda_b] \Phi(q),
\eez
so that
\beq
\Upsilon(q,\lambda,\Pi,\overline{\Pi}) =    \exp    [-\overline{\Pi}_a
M_b{}^a \Pi^b]  \exp [\lambda^a M_a{}^b \Lambda_b] \Phi(q).
\l{razumov-1}
\eeq
Substituting this expression to the inner product \r{12-1}, one finds
\bez
(\Upsilon,\Upsilon) =   \int   dq   \prod_{a=1}^M   d\mu^a   \Phi^*(q)
e^{- 2i\mu^a M_a{}^b  \Lambda_b}    \Phi(q)     \int     \prod_{a=1}^M
d\overline{\Pi}_a d\Pi^a \exp [-2\overline{\Pi}_a M_b{}^a \Pi^b].
\eez
Integration over Grassmannian variables gives us the factor $\det  2M$
which  is  involved  to  the  integration  measure  after substitution
$2\mu^a M_a{}^b = \tilde{\mu}^b$. We obtain that
\beq
(\Upsilon,\Upsilon) =         (\Phi,         \prod_{a=1}^M        2\pi
\delta(\hat{\Lambda}_a) \Phi).
\l{5a-2}
\eeq
This formula  can  be  valid for the case of the continuous spectrum of
constraints. For the discrete spectrum case,  we see  that  there  are
internal topological  problems  in  BRST-BFV approach (cf.\c{DSS}).  A
possible resolution of them is to  modify  formula  \r{12-1}  for  the
BRST-BFV inner product by integration over $\mu^1,...,\mu^M$ belonging
to some domain rather than over $\mu \in {\bf R}^M$.
Formula \r{5a-2} coincides with  the  inner  product  in  the
refined algebraic   quantization   approach   \r{3-1},  provided  that
$\Upsilon$ is taken to the gauge \r{9-1} and auxiliary state $\Phi$ is
\beq
\Phi(q) = \Upsilon(q,0,0,0).
\l{6-2}
\eeq
Let $\tilde{\Upsilon}$  be a physical state which does not satisfy the
gauge conditions \r{9-1}. This means that
\beq
\tilde{\Upsilon} = {\Upsilon} + \Omega X,
\l{7-2}
\eeq
while ${\Upsilon}$ obeys gauge condition \r{9-1}. Consider the function
\bez
\tilde{\Phi}(q) = \tilde{\Upsilon}(q,0,0,0)
\eez
and investigate the relationship between $\Phi$ and $\tilde{\Phi}$. Let
\bez
X =   X_{00}  (q,\lambda)  +  X_{01}^a(q,\lambda)  \overline{\Pi}_a  +
X_{10,a}(q,\lambda) \Pi^a + ...
\eez
Then
\beq
\tilde{\Phi}(q) - \Phi(q) = \hat{\Lambda}_a X^a_{01}(q,0),
\l{8-2}
\eeq
so that  $\tilde{\Phi}$  and  $\Phi$  are gauge-equivalent in sense of
\r{4-1}. Thus, if $\tilde{\Upsilon}$ is an arbitrary physical state in
the BRST-BFV-approach,  formula  \r{6-2}  gives us a representative of
class of equivalence of auxiliary states.  Thus,  for abelian case the
correspondence between  BFV  and refined algebraic quantization states
is found.  If one used a coordinate  representation  with  respect  to
ghosts and  antighosts  instead  of momenta representation,  $\Phi(q)$
would be a component of the BFV  wave  function  with  maximal  number
ghosts and antighosts.

To obtain a correspondence between Dirac and BFV states,  consider the
integral
\beq
\Psi(q) =    \int    \prod_a    d\mu^a    d\overline{\Pi}_a     d\Pi^a
\Upsilon(q,-i\mu, \Pi, \overline{\Pi}).
\l{Dirac-1}
\eeq
For the  gauge  \r{14-1},  $\Upsilon$  has  the  form   \r{razumov-1}.
Substituting it to formula \r{Dirac-1}, we find
\bez
\Psi(q) = \prod_{a=1}^M 2\pi \delta (\hat{\Lambda}_a) \Phi(q).
\eez
This formula coincides with \r{5-1}. Therefore, to find the Dirac wave
function, one should take the B-state to the gauge  \r{14-1}  and  use
eq.\r{Dirac-1}. However,  formula  \r{Dirac-1}  is valid for arbitrary
physical state. Namely,
\bez
\int    \prod_a    d\mu^a    d\overline{\Pi}_a     d\Pi^a
(\Omega X)(q,-i\mu, \Pi, \overline{\Pi})
=
\int    \prod_a    d\mu^a    d\overline{\Pi}_a     d\Pi^a
(\frac{1}{i} \frac{\partial}{\partial\mu^a} \Pi^a +
\frac{\partial}{\partial \overline{\Pi}_a}\hat{\Lambda}_a)
X (q,-i\mu, \Pi, \overline{\Pi}) = 0.
\eez
We have found that for abelian case the Dirac and BRST-BFV states  are
related with  the  help  of  formula  \r{Dirac-1}.  In  the coordinate
representation with respect to  ghosts,  formula  \r{Dirac-1}  can  be
viewed as  a component of the B-function with minimal number of ghosts
and antighosts.

\subsection{Closed-algebra case}

Let us generalize formulas \r{6-2} and \r{Dirac-1} to  the  nonabelian
case and  check  them  for  the Batalin-Marnelius prescription for
the inner product.

Suppose that the B-state $\Upsilon$ is taken  to  the  gauge  \r{9-1}.
This means that $\Upsilon$ is $\lambda,\Pi,\overline{\Pi}$-independent,
\beq
\Upsilon = \Phi(q)
\l{9-2}
\eeq
provided that the ghost number of  $\Upsilon$  is  zero  (eq.\r{4-2}).
Since
\beq
[\Omega,\rho]_+ = -\lambda^a \Lambda_a + \frac{i}{2} \lambda^a f^b_{ab}
- i\lambda^a \overline{\Pi}_b C^c f^b_{ac} - \overline{\Pi}_a \Pi^a,
\l{12*-1}
\eeq
for the simplest abelian case one has
\beq
(e^{t[\Omega,\rho]_+}\Upsilon) (q,\lambda,  \Pi,   \overline{\Pi})   =
e^{-t\lambda^a\Lambda_a} \Phi(q) e^{-t\overline{\Pi}_a\Pi^a},
\l{12+-1}
\eeq
so that
\bez
(\Upsilon, e^{t[\Omega,\rho]_+} \Upsilon) = \int dq
\prod_{a=1}^M     d\mu^a
d\overline{\Pi}_a d\Pi^a
\Phi^*(q) e^{it\mu^a\Lambda_a} e^{-t\overline{\Pi}_a \Pi^a} \Phi(q).
\eez
Integration over ghost variables gives us $t^M$, so that
\beq
(\Upsilon, e^{t[\Omega,\rho]_+} \Upsilon) =
(\Phi, \prod_{a=1}^M 2\pi \delta(\hat{\Lambda}_a) \Phi).
\l{10-2}
\eeq
We see that inner products \r{10-1}  and  \r{3-1}  coincide,  provided
that correspondence  between  $\Phi$  and  $\Upsilon$  is  of the form
\r{6-2}. Thus,  formula \r{6-2} is justified  even  if  the  Marnelius
inner product is introduced in the abelian theory.
We also   see   that   for   the   case   of   discrete   spectrum  of
$\hat{\Lambda}_a$ topological problems arise in this  version  of  the
BRST-BFV approach as well.

Formula \r{Dirac-1}  contains  factors  $0\times \infty$ and should be
then regularized as
\bez
\Psi(q) =    \int    \prod_a    d\mu^a    d\overline{\Pi}_a     d\Pi^a
(e^{t[\Omega,\rho]_+} \Upsilon) (q,-i\mu, \Pi, \overline{\Pi}).
\eez

Evaluating this integral,  we also  obtain  formula  \r{5-1}.  Formula
\r{Dirac-1} is justified.

Consider now the nonabelian case. Let us look for
the wave function $e^{t[\Omega,\rho]_+}\Upsilon$ in the following form:
\bez
(e^{t[\Omega,\rho]_+}\Upsilon)(q,\lambda,\Pi,\overline{\Pi})         =
e^{- t\lambda^a \hat{\Lambda}_a} \Phi(q)  e^{\overline{\Pi}_a  B^a{}_b
(\lambda,t) \Pi^b}
\eez
where $\hat{\Lambda}_a$ is of the form \r{2-2}
Making use of the relation
$\frac{d}{dt}(e^{t[\Omega,\rho]_+} \Upsilon) = [\Omega,\rho]_+
e^{t[\Omega,\rho]_+} \Upsilon$, we find the following equation for the
matrix $B$,
\bez
\dot{B}^b{}_d = - i\lambda^a f^b_{ac} B^c{}_d - 1,
\eez
so that
\bez
B(\lambda,t) = -\int_0^t d\tau Ad\{\exp(- \tau \lambda^aL_a\} \}.
\eez
Therefore,
\beq
(\Upsilon, e^{t[\Omega,\rho]_+}  \Upsilon)  =  \int  dq \prod_a d\mu^a
d\overline{\Pi}_a d\Pi^a  \Phi^*(q)    e^{it\mu^a    (\check{\Lambda}_a    -
\frac{i}{2} f^b_{ab})}\Phi(q)
e^{-\overline{\Pi}_a
\int_0^t d\tau
(Ad \{ \exp(i\tau \mu^cL_c)\})^a_b \Pi^b}.
\l{15-1}
\eeq
Integration over fermionic variables gives us the group measure
\bez
dg = \det
\int_0^t d\tau (Ad \{ \exp(i\tau \mu^cL_c)\}) \prod_{a=1}^{M} d\mu^a,
\qquad g = \exp(it\mu^cL_c)
\eez
It happens that it coincides with the  right-invariant  Haar  measure
which has the form (see,  for example, \c{group}) $d_Rg = d\mu J(\mu)$
with $J(\mu) = \det \frac{\delta \rho}{\delta \mu}$ for
\bez
\exp(i(\mu^a+\delta\mu^a)L_a) =        \exp(i\delta         \rho^aL_a)
\exp(i\mu^aL_a).
\eez
Without loss of generality, consider the case $t=1$. One finds
\beq
\delta\rho^aL_a = \int_0^1 d\alpha e^{i\alpha \mu^aL_a} \delta\mu^b L_b
e^{-i\alpha \mu^aL_a}  =  \int_0^1 d\alpha (Ad\{e^{i\alpha \mu^cL_c}\}
\delta\mu)^a L_a,
\l{15a-1}
\eeq
so that $dg=d_Rg$. The multiplicator $e^{t\frac{1}{2} \mu^af_{ab}^b}$
can be presented as $(det Ad\{g\})^{-1/2}$.  The inner product  \r{15-1}
occurs to coincide with \r{16-1},  provided  that  the  correspondence
between $\Phi$  and  $\Upsilon$  is  of  the form \r{6-1}.  Under gauge
transformation \r{7-2} of $\Upsilon$ the  auxiliary  state  $\Phi$  is
transformed according  to  eq.\r{8-2}.  This is a gauge transformation
\r{4-1}. Thus,  formula \r{6-2} is checked for the nonabelian case  as
well.

Analogously to the abelian case,  one can propose that each  point  of
the gauge group should be taken into account once. This means that the
inner product \r{12-1} should be in general modified: integration over
$\mu$ should be performed over some domain only.

The Dirac wave function \r{17a-1} can be also presented
via the integral over ghost momenta and
Lagrange multipliers,
\beq
\Psi(q) =       \prod_a      d\mu^a      d\overline{\Pi}_a      d\Pi^a
(e^{t[\Omega;\rho]_+} \Phi)(q, -i\mu, \Pi, \overline{\Pi}).
\l{11-2}
\eeq
Since states  $\Phi$  and  $e^{t[\Omega;\rho]_+}  \Phi$  are  formally
BFV-equivalent, one can notice that eq.\r{11-2} is formally equivalent
to \r{Dirac-1}.  Namely, gauge-equivalent BFV-states give us identical
Dirac wave functions,  since the BRST-BFV charge can be written  as  a
full derivative,
\bez
\Omega =  (\check{\Lambda}_a
+ \frac{i}{2} f^b_{ab}) \frac{\partial}{\partial
\overline{\Pi}_a} -  \frac{i}{2}   f^a_{bc}
\frac{\partial}{\partial \overline{\Pi}_b}
\frac{\partial}{\partial \overline{\Pi}_c}
\overline{\Pi}_a - \frac{\partial}{\partial \lambda^a} \Pi^a,
\eez
so that the integral of $\Omega X$ over $\mu$, $\overline{\Pi}$, $\Pi$
vanishes. Furthermore, it follows from the property
\bez
\int \prod_a d\mu^a d\overline{\Pi}_a d\Pi^a (\Omega  \overline{\Pi}_a
\Upsilon)(q,-i\mu,\Pi, \overline{\Pi})
\eez
and relation  $\Omega  \Upsilon=0$  that eq.\r{18-1} is indeed satisfied
for definition \r{Dirac-1}.

Thus, the   formal  relationship between Dirac and BFV states is
obtained.  We also see that integration over $\mu$ should be performed
carefully   due   to  topological  problems.

\section{A proposal  for  the inner product and its properties for the
open-algebra case}

\subsection{Prescription for the inner product}

To develop the method of refined algebraic quantization for  the  case
of nontrivial structure functions, let us suppose that formula \r{6-2}
for the correspondence between BFV and  refined  algebraic  states  is
valid. Then  we will write down the Batalin-Marnelius prescription for
BFV inner product and find the operator  $\eta$  entering  to  formula
\r{4}.

The quantum constrained system is specified by the B-charge \r{7}. Let
$\Upsilon= \Phi(q)$ satisfy conditions \r{9-1}.  Calculate  the  inner
product \r{10-1}. One has
\bez
[\Omega,\rho]_+ = - \overline{\Pi}_a \Pi^a - \lambda^a \hat{\Omega}_a
\eez
with
\bez
\hat{\Omega}_a =   \Omega_a  (\overline{\Pi},C)  =  [\overline{\Pi}_a,
\Omega]_+ = \hat{\Lambda}_a + ... + n
\Omega^n{}^{b_1...b_{n-1}}_{a_1...a_{n-1}a}
\overline{\Pi}_{b_1}
...
\overline{\Pi}_{b_{n-1}}
C^{a_1}... C^{a_{n-1}} + ...
\eez
We obtain then that
\beq
(\Phi, e^{t[\Omega,\rho]_+}\Phi) =
\int dq \Phi^*(q) \prod_{a=1}^M d\mu^a d\overline{\Pi}_a d\Pi^a
e^{-t\overline{\Pi}_a \Pi^a + it\mu_a \hat{\Omega}_a} \Phi(q)
\l{15}
\eeq
with $\hat{\Omega}_a  =
\Omega_a  (\overline{\Pi},  \partial/\partial \overline{\Pi})$.
Formula \r{15} is of the type \r{4} with
\beq
\eta = \int \prod_{a=1}^M d\mu^a d\overline{\Pi}_a d\Pi^a
e^{-\overline{\Pi}_a \Pi^a + i\mu_a \hat{\Omega}_a
(\overline{\Pi},  \partial/\partial \overline{\Pi})} 1.
\l{16}
\eeq
Here $\Pi^a$ and $\mu_a$ are rescaled in $t$ times.

One should  take  into account the topological problems analogously to
the closed-algebra case: integration may be performed not over
all values of $\mu$ but over $\mu$ belonging to some domain.

\subsection{Properties of the inner product}

Let us investigate properties of the operator $\eta$ \r{16}.
First of all,  check that $\eta^+ = \eta$, so that formula \r{4} gives
us real values. One has
\bez
(\Phi, \eta  \Phi)  =  \int  \prod_{a=1}^M  d\mu^a  (\Phi,  \exp[\Pi^a
\overline{\Pi}_a + i \mu_a \hat{\Omega}_a] \Phi)
\eez
and
\bez
(\Phi, \eta \Phi)^* = \int  \prod_{a=1}^M  d\mu^a  (\Phi,
\exp[\Pi^a \overline{\Pi}_a - i \mu_a \hat{\Omega}_a^+] \Phi)
\eez
After change  of  variables $\mu_a \to - \mu_a$ and using the property
$\hat{\Omega}_a^+ = \hat{\Omega}_a$ being a corollary of the relations
$\overline{\Pi}_a^+ =  \overline{\Pi}_a$  and $\Omega^+ = \Omega$,  we
find $\eta^+ = \eta$.

Let us check relation \r{5}. One has
\beq
\eta \hat{\Lambda}_b Y^b(q) = \int
\prod_{a=1}^M d\mu^a d\overline{\Pi}_a d\Pi^a
\exp[\Pi^a \overline{\Pi}_a   +   i   \mu_a   \hat{\Omega}_a]   \Omega
\overline{\Pi}_b Y^b(q).
\l{17}
\eeq
Since $\Omega^2=\Omega$,  the operators $\Omega$ and  $[\Omega,\rho]_+$
commute:
\beq
\Omega [\Omega,\rho]_+ = \Omega \rho \Omega = [\Omega,\rho]_+ \Omega,
\l{17a}
\eeq
so that
\bez
e^{\Pi^a \overline{\Pi}_a + i\mu_a \hat{\Omega}_a} \Omega =
\Omega e^{\Pi^a \overline{\Pi}_a + i\mu_a \hat{\Omega}_a}.
\eez
Formula \r{17} transforms then to
\beq
\eta \hat{\Lambda}_b Y^b(q) = \int
\prod_{a=1}^M d\mu^a d\overline{\Pi}_a d\Pi^a
\Omega^+
\exp[\Pi^a \overline{\Pi}_a   +   i   \mu_a   \hat{\Omega}_a]
\overline{\Pi}_b Y^b(q)
\l{18}
\eeq
since $\Omega = \Omega^+$. The operator $\Omega^+$ can be presented in
representation \r{14a} as
\beq
\Omega^+ =      \frac{1}{i} \frac{\partial}{\partial \mu_a}   \Pi^a   +
\frac{\partial}{\partial \overline{\Pi}_a}   \hat{\Lambda}_a^+   +
...  +
(\Omega^n{}^{b_1...b_{n-1}}_{a_1...a_n})^+
\frac{\partial}{\partial \overline{\Pi}_{a_n}} ...
\frac{\partial}{\partial \overline{\Pi}_{a_1}}
\overline{\Pi}_{b_{n-1}}
... \overline{\Pi}_{b_1} + ...
\l{BRST}
\eeq
Integral \r{18} then vanishes  as  an  integral  of  full  derivative.
Formula \r{5} is checked.
Thus, formula \r{16} obeys the  desired  properties  of  the  operator
$\eta$ entering   to  the  inner  product.  However,
the problem of positive definiteness of the inner product  remains  to
be investigated.

\subsection{Correspondence of  states  for  the  case  of   nontrivial
structure functions}

Let us  show  that  correspondence of BFV,  auxiliary and Dirac states
given by eqs.\r{6-2} and \r{Dirac-1} remains valid  for  the  case  of
nontrivial structure functions.

First, notice that the auxiliary state $(\Omega X)(q,0,0,0)$ is of the
form
\bez
(\Omega X)(q,0,0,0)    =    \hat{\Lambda}_a   \frac{\partial}{\partial
\overline{\Pi}_a}|_{\overline{\Pi}=\Pi=0, \lambda = 0} X
\eez
and is equivalent then to zero.  Thus,  refined algebraic quantization
state $\Upsilon  (q,0,0,0)  =  (\Omega  X)(q,0,0,0)  +   \Phi(q)$   is
equivalent to $\Phi(q)$. Formula \r{6-2} is justified.

The Dirac wave function \r{6} can be presented as
\bez
\Psi(q) =   \int   \prod_{a=1}^M   d\mu^a   d\overline{\Pi}_a   d\Pi^a
(e^{[\Omega,\rho]_+} \Phi)(q, -i\mu, \Pi, \overline{\Pi}).
\eez
To check  formula  \r{Dirac-1},  it  is  sufficient  to  justify  that
equivalent BFV states \r{10} give
equal Dirac  wave functions \r{Dirac-1}.  However,  it follows directly
from \r{BRST} that
\bez
\int   \prod_{a=1}^M   d\mu^a   d\overline{\Pi}_a   d\Pi^a
(\Omega^+ X) (q, -i\mu, \Pi, \overline{\Pi}) = 0
\eez
since integrals  of  full  derivatives  vanish.  Formula \r{Dirac-1} is
obtained.

To check relation \r{1}, use the property
\bez
\int   \prod_{a=1}^M   d\mu^a   d\overline{\Pi}_a   d\Pi^a
(\Omega^+ \overline{\Pi}_a \Upsilon)
(q, -i\mu, \Pi, \overline{\Pi}) = 0.
\eez
Since $\Omega^+\Upsilon  =  \Omega\Upsilon =0$,  one can rewrite it as
follows,
\beq
\int   \prod_{a=1}^M   d\mu^a   d\overline{\Pi}_a   d\Pi^a
([\Omega^+, \overline{\Pi}_a]_+ \Upsilon)
(q, -i\mu, \Pi, \overline{\Pi}) = 0.
\l{i1}
\eeq
The anticommutator has the form
\bez
[\Omega^+;\overline{\Pi}_a]  =
\hat{\Lambda}_a^+   +
...  +
n (\Omega^n{}^{b_1...b_{n-1}}_{a_1...a_{n-1}a})^+
\frac{\partial}{\partial \overline{\Pi}_{a_{n-1}}} ...
\frac{\partial}{\partial \overline{\Pi}_{a_1}}
\overline{\Pi}_{b_{n-1}}
... \overline{\Pi}_{b_1} + ...
\eez
It contains full derivatives, except for the term $\hat{\Lambda}_a^+$.
Thus, eq.\r{i1} can be presented as
\bez
\hat{\Lambda}_a^+
\int   \prod_{a=1}^M   d\mu^a   d\overline{\Pi}_a   d\Pi^a
\Upsilon (q, -i\mu, \Pi, \overline{\Pi}) = 0.
\eez
Eq.\r{1} is obtained.

Thus, the   {\it  formal  }  relationship  between  refined  algebraic
quantization,  Dirac and BFV  states  is  found.  However,  there  are
topology  problems  which  may lead to integration over some domain in
\r{Dirac-1}. They are to be investigated in future.

\section{Correspondence of observables}

Let us  consider  the  properties  of quantum observables in different
quantization approaches.

In the BRST-BFV approach, observables are viewed as series
\beq
H_B = H + ... +
H^n{}^{b_1...b_{n}}_{a_1...a_n} \overline{\Pi}_{b_1}               ...
\overline{\Pi}_{b_n} C^{a_1} ... C^{a_n} + ...
\l{12-2}
\eeq
The operator  coefficient  functions  $H^n{}^{b_1...b_{n}}_{a_1...a_n}
(\hat{p},\hat{q})$
are chosen in such a way that
\beq
H_B^+ = H_B, \qquad [\Omega, H_B] = 0.
\l{ham}
\eeq
These properties  provide  that physical states \r{9} are taken by the
operator $H$  to  physical,  while  equivalent  states  are  taken  to
equivalent; the inner product is conserved under evolution.

One has:
\bez
\Omega H_B  =  C^c  \hat{\Lambda}_c H + \hat{\Lambda}_b H_a^{1b} C^a +
..., \qquad H_B \Omega = HC^c \hat{\Lambda}_c + ...
\eez
where $...$ are terms with ghost momenta.  Therefore,  $H$ should obey
the following property:
\bez
[H;\hat{\Lambda}_a] = \hat{\Lambda}_b H_a^{1b}
\eez
for some operators $H_a^{1b}$. We have obtained relation \r{3}.

Since $(H_B\Upsilon)(q,0,0,0) = H\Upsilon(q,0,0,0)$,
it is  the  operator $H$ that corresponds to the B-observable \r{12-2}
in the refined algebraic quantization approach.  An important  feature
of the physical observable is that the corresponding evolution operator
$e^{-iHt}$ should be unitary with respect to the inner product  \r{4}.
This means that
\bez
(e^{-iHt})^+ \eta e^{-iHt} = \eta.
\eez
or
\beq
H^+\eta = \eta H.
\l{6a}
\eeq
This property is to be checked.

Let $\Phi$ be an auxiliary state  corresponding  to  the  Dirac  state
$\Psi=\eta\Phi$.
The observable $H$  takes  it  to  $H\Phi$.  This
corresponds to the Dirac state
\bez
\eta H\Phi = H^+ \eta \Phi = H^+\Psi.
\eez
Therefore, it is the operator $H^+$ that corresponds to the observable
$H$ in  the  Dirac  approach,  while  $\exp(-iH^+t)$  is  an evolution
operator. This is also a corollary of \r{Dirac-1}
since
\bez
\int \prod_{a=1}^M     d\mu_a     d\overline{\Pi}_a    d\Pi^a    H_B^+
\Upsilon(q,-i\mu,\Pi,\overline{\Pi})
= H^+ \int \prod_{a=1}^M     d\mu_a     d\overline{\Pi}_a    d\Pi^a
\Upsilon(q,-i\mu,\Pi,\overline{\Pi})
\eez
because integral of full derivative vanishes.

\subsection{"Closed-algebra" case}

Let us   check   formula   \r{6a}.   First   of   all,   consider  the
"closed-algebra" case with constant  operators  $H^{1b}_a  =  iR_a^b$,
$R_a^b = const$.  The higher-order terms of expression \r{12-2} vanish
then \c{Henneaux}, so that
\beq
H_B = H + i\overline{\Pi}_b R_c^b C^c.
\l{22-1}
\eeq
Eq.\r{6a} to be checked can be rewritten as
\beq
\int d_Rg H^+ e^{i\mu^a\hat{\Lambda}_a}
= \int d_Rg e^{i\mu^a\hat{\Lambda}_a} H.
\l{23-1}
\eeq
One has
\bez
e^{i\mu^a\hat{\Lambda}_a} H  e^{-i\mu^a\hat{\Lambda}_a}  =   H   +
\int_0^1 d\alpha e^{i\alpha\mu^a\hat{\Lambda}_a}
[i\mu^a\hat{\Lambda}_a; H]
e^{-i\mu^a\hat{\Lambda}_a} =
H +
\int_0^1 d\alpha e^{i\alpha\mu^a\hat{\Lambda}_a}
\mu^a R_a^b \hat{\Lambda}_b
e^{-i\mu^a\hat{\Lambda}_a}.
\eez
Since the commutation relations between generators $\hat{\Lambda}_a$
coincide with    \r{1-1},    $[\hat{\Lambda}_a,\hat{\Lambda}_b]    =
if^c_{ab} \hat{\Lambda}_c$, it follows from eq.\r{15a-1} that
\bez
e^{i\mu^a\hat{\Lambda}_a} H  e^{-i\mu^a\hat{\Lambda}_a}  =   H   +
\frac{1}{i} \frac{d}{d\tau}|_{\tau=0}
e^{i(\mu^a + \tau \mu^b R_b^a) \hat{\Lambda}_a}
e^{-i \mu^a \hat{\Lambda}_a}.
\eez
Eq. \r{23-1} is taken then to the form
$\int d_Rg (H^+ - H) e^{i\mu^a\hat{\Lambda}_a}
= \int d_Rg  \frac{1}{i}  \frac{d}{d\tau}|_{\tau=0}
e^{i(\mu^a + \tau \mu^b R_b^a) \hat{\Lambda}_a}$,
so that
\beq
H=H^+ - iR_b^b
\l{24-1}
\eeq
Condition \r{24-1}  is  a relationship between observables $H$ and $H^+$
in the projection operator and Dirac approaches.  We see that this  is
in agreement with the condition $H_B^+=H_B$.

\subsection{General case}

Let us verify formula \r{6a} for general case.
One has
\beq
\eta H \Phi(q) = \int
\prod_{a=1}^M d\mu_a d\overline{\Pi}_a d\Pi^a
\exp[\Pi^a \overline{\Pi}_a   +   i   \mu_a   \hat{\Omega}_a]
H_B \Phi(q),
\l{19}
\eeq
while
\beq
H^+ \eta  \Phi(q) = \int
\prod_{a=1}^M d\mu_a d\overline{\Pi}_a d\Pi^a
H_B^+
\exp[\Pi^a \overline{\Pi}_a   +   i   \mu_a   \hat{\Omega}_a]
\Phi(q).
\l{20}
\eeq
Here we  have  taken  into account that $C^a \Phi(q) = 0$ and that the
integral of full  derivative  vanishes.  Consider  the  difference  of
eqs.\r{19}, \r{20}. Let us make use of the following relation,
\bez
H_B^+ e^{[\Omega,\rho]_+} -  e^{[\Omega,\rho]_+} H_B = \int_0^1 d\tau
e^{\tau[\Omega,\rho]_+} [[\Omega,\rho]_+,       H_B]       e^{(1-\tau)
[\Omega,\rho]_+},
\eez
since $H_B^+    =    H_B$.    Moreover,    $[[\Omega,\rho]_+,H_B]    =
[\Omega,[H_B,\rho]]_+$. It follows from eq.\r{17a} that
\bez
H_B^+ e^{[\Omega;\rho]_+} - e^{[\Omega;\rho]_+} H_B = [\Omega;A]_+
\eez
with
\bez
A= \int_0^1 d\tau
e^{\tau[\Omega,\rho]_+} [H_B; \rho]       e^{(1-\tau)
[\Omega,\rho]_+},
\eez
Therefore, the difference between formulas \r{19} and \r{20} reads
\bez
(H^+\eta - \eta H)\Phi(q) = \int
\prod_{a=1}^M d\mu^a d\overline{\Pi}_a d\Pi^a
[\Omega^+A + A\Omega] \Phi(q).
\eez
This integral vanishes since $\Omega \Phi(q) = 0$ and  an  integral  of
full derivative is zero. Thus, relation \r{6a} is satisfied.

\section{A simple example of system with nontrivial structure functions}

Consider a simple  example  of  a  system  with  structure  functions.
Investigate the   model   with   3  degrees  of  freedom  $(p_i,q^i)$,
$i=\overline{1,3}$ and 2 classical constraints
\beq
\Lambda_1 = a(q^2,q^3) p_1, \qquad \Lambda_2 = p_2.
\l{20a}
\eeq
Since $\{\Lambda_1,   \Lambda_2\}   =   \partial_2   \log   a(q^2,q^3)
\Lambda_1$, the constraints forms an algebra with structure functions.
Let us  look for the B-charge in the form \r{7}.  In classical theory,
it should be written as
\beq
\Omega = -i\pi_1 \Pi^1 - i \pi_2 \Pi^2  +  p_1  a  C^1  +  p_2  C^2  +
(\alpha_1 \overline{\Pi}_1 + \alpha_2 \overline{\Pi}_2) C^1C^2
\l{21}
\eeq
for some functions $\alpha_a(p,q)$.  The property $\{\Omega,\Omega\} =
0$ means that
\bez
p_1 a \alpha_1 + p_2\alpha_2 = [p_1a; p_2],
\eez
so that
\bez
\alpha_1 = - i \partial_2 \log a; \qquad \alpha_2 = 0.
\eez
We see that classically
\bez
\Omega = -i\pi_1 \Pi^1 - i\pi_2\Pi^2 + p_1aC^1 + p_2C^2 - i \partial_2
\log a \overline{\Pi}_1 C^1C^2.
\eez
To quantize the B-charge,  one should choose the operator ordering. If
the $\overline{\Pi}$-operators  were  put  to the left with respect to
$C$-operators, the quantum B-charge would be not  Hermitian.  To  obey
the condition $\Omega^+=\Omega$, let us use the Weyl quantization
\beq
\Omega =   -i\pi_1\Pi^1   -   i   \pi_2\Pi^2   +   p_1aC^1  +  (p_2  -
i\overline{\Pi}_1 \partial_2 \log a C^1 + \frac{i}{2} \partial_2  \log
a)C^2
\l{22}
\eeq
It is   remarkable   that   in   quantum   theory    the    constraint
$\hat{\Lambda}_2$ should  be  modified  with  respect to the classical
theory \r{20a}; it follows from eq.\r{7} that
\bez
\hat{\Lambda}_2 = p_2 + \frac{i}{2} \partial_2 \log a,
\eez
so that the operator $\hat{\Lambda}_2$ becomes formally non-Hermitian.
This feature  of  quantum  constraints  is  known  from  the theory of
constrained systems with nonunimodular closed algebra \c{KS,M2}.

Let us evaluate the inner product \r{15}. Consider the wave function
\beq
\Upsilon^t(q, \overline{\Pi},  \Pi) = e^{- t\overline{\Pi}_a \Pi^a +  it
\mu_a \hat{\Omega}_a} \Phi(q).
\l{23}
\eeq
Since
\bez
\hat{\Omega}_1 = p_1a + i\overline{\Pi}_1 \partial_2 \log a C^2,
\qquad \hat{\Omega}_2  = p_2 - i\overline{\Pi}_1 \partial_2 \log a C^1
+ \frac{i}{2} \partial_2 \log a,
\eez
the state \r{23} obeys the following Cauchy problem
\beq
\frac{\partial}{\partial t}\Upsilon^t   =   [- \overline{\Pi}_1\Pi^1   -
\overline{\Pi}_2 \Pi^2  +  a  \mu_1  \partial_1  +  \mu_2 \partial_2 -
\frac{\mu_2}{2} \partial_2 \log a - \mu_1 \overline{\Pi}_1  \partial_2
\log a     \frac{\partial}{\partial    \overline{\Pi}_2}    +    \mu_2
\overline{\Pi}_1 \partial_2    \log     a     \frac{\partial}{\partial
\overline{\Pi}_1}] \Upsilon^t,
\l{24}
\eeq
\bez
\Upsilon^0 = \Phi(q)
\eez
Since eq.\r{24} is a first-order partial differential equation, it can
be solved by the characteristic method.  The solution is looked for in
the following form
\bez
\Upsilon^t(Q^t,\tilde{\Pi}^t,\Pi) =         \exp[\int_0^t        d\tau
[- \tilde{\Pi}_1^{\tau} \Pi^1    -    \tilde{\Pi}_2^{\tau}    \Pi^2    -
\frac{\mu_2}{2} \partial_2            \log a(Q^{\tau})]               ]
\Upsilon^0(Q^0,\tilde{\Pi}^0,\Pi),
\eez
where the  functions  $Q^t$,  $\tilde{\Pi}^t$  satisfy  the  following
ordinary differential equations,
\bez
\dot{Q}_1^t = - a(Q_2,Q_3)\mu_1;  \qquad \dot{Q}_2^t = -\mu_2,  \qquad
\dot{Q}_3^t = 0,
\eez
\bez
\frac{d}{dt}\tilde{\Pi}_2^t =   \mu_1  \tilde{\Pi}_1  \partial_2  \log
a(Q_2,Q_3),
\qquad
\frac{d}{dt}\tilde{\Pi}_1^t = -  \mu_2  \tilde{\Pi}_1  \partial_2  \log
a(Q_2,Q_3),
\eez
so that the classical characteristic trajectory is
\bez
Q_3^t = Q_3^0, \qquad
Q_2^t = Q_2^0 - \mu_2 t, \qquad
Q_1^t = Q_1^0 - \int_0^t d\tau a(Q_2^0-\mu_2\tau, Q_3^0) \mu_1,
\eez
\bez
\tilde{\Pi}_1^t =   \frac{a(Q_2^0  -  \mu_2t,  Q_3^0)}{a(Q_2^0,Q_3^0)}
\tilde{\Pi}_1^0, \qquad
\tilde{\Pi}_2^t =  \tilde{\Pi}_2^0  +  \frac{1}{a(Q_2^0,Q_3^0)}  \mu_1
\tilde{\Pi}_1^0 \int_0^t d\tau \partial_2 a(Q_2^0-\mu_2\tau,Q_3^0).
\eez
Combining all factors,  one finds the solution the Cauchy  problem
\r{24},
\bez
\Upsilon^t(x,\overline{\Pi},\Pi) =
\sqrt{\frac{a(x_2,x_3)}{a(x_2+\mu_2t, x_3)}}   \exp[- \int_0^t
d\tau \frac{a(x_2+\mu_2t, x_3)}{a(x_2, x_3)} \overline{\Pi}_1 \Pi^1
- t\overline{\Pi}_2 \Pi^2]
\eez
\beq
\times
\exp[  \int_0^t d\tau \tau \mu_1
\frac{\partial_2 \log                    a(x_2+\mu_2\tau)}{a(x_2,x_3)}
\overline{\Pi}_1\Pi^2]
\Phi(x_1 + \int_0^t d\tau a(x_2+\mu_2\tau,x_3)\mu_1, x_2+\mu_2t, x_3)
\l{25}
\eeq
One can also check by the direct computations that  expression  \r{25}
really satisfies  the Cauchy problem \r{24}.  The inner product \r{15}
reads
\beq
\int dx  \Phi^*(x)  \prod_{a=1}^M  d\mu_a   d\overline{\Pi}_a d\Pi^a
\Upsilon^t(x,\overline{\Pi},\Pi).
\l{26}
\eeq
Integration over ghost variables gives us the multiplier
\bez
t\int_0^t d\tau a(x_2+ \mu_2\tau, x_3) \frac{1}{a(x_2,x_3)}.
\eez
After rescaling of variables $\mu$
\bez
\xi_1 = \int_0^t d\tau a(x_2 + \mu_2\tau,x_3) \mu_1, \qquad
\xi_2 = t\mu_2
\eez
one finds that the integral \r{26} takes a simple form
\bez
\int dx_1dx_2dx_3  d\xi_1  d\xi_2  \Phi^*(x_1,x_2,x_3) \frac{1}{\sqrt{
a(x_2+\xi_2,x_3) a(x_2,x_3) }} \Phi(x_1+\xi_1, x_2+ \xi_2,x_3)
\eez
We see that the bilinear form $\eta$ can be defined as
\bez
(\Phi,\eta \Phi)      =      \int      dx_3       |\int       dx_1dx_2
\frac{\Phi(x_1,x_2,x_3)}{\sqrt{a(x_2,x_3)}}|^2,
\eez
so that  the correspondence between the Dirac wave function $\Psi$ and
the auxiliary state $\Phi$ is
\bez
\Psi(x_1,x_2,x_3) = \frac{1}{\sqrt{a(x_2,x_3)}} \int dy_1 dy_2
\frac{\Phi(y_1,y_2,x_3)}{\sqrt{a(y_2,x_3)}}.
\eez
It obeys the constraints
\bez
a(x_2,x_3) p_1 \Psi \equiv \hat{\Lambda}_1^+ \Psi = 0, \qquad
\frac{1}{\sqrt{a(x_2,x_3)}} p_2    \sqrt{a(x_2,x_3)}    \Psi    \equiv
\hat{\Lambda}_2^+ \Psi = 0.
\eez
while the gauge transformation of $\Phi$ is
\bez
\Phi \to  \Phi + \sqrt{a(x_2,x_3)} p_2 \frac{1}{\sqrt{a(x_2,x_3)}} Y^2
+ a(x_2,x_3) p_1 Y^1
\eez
for some functions $Y^1$ and $Y^2$.  We see  that  all  properties  of
$\eta$ (including positive definiteness)
are indeed satisfied in this example.

\section{Discussion}

Thus, the correspondence between states and observables  in  BRST-BFV,
Dirac and  refined  algebraic  quantizations  is found.  For different
versions of BRST-BFV approach,  the relations \r{6-2} and  \r{Dirac-1}
for auxiliary and Dirac state vectors are found.  However, it has been
noticed that there are internal problems in  the  BRST-BFV  formalism.
They arise   when   the   constraint   gauge  group  is  topologically
nontrivial. For  simple  examples,   the   topological   problems   of
B-approach can be resolved by integrating over $\mu$ belonging to some
domain in the inner product formula \r{12-1}.  One can hope that  this
prescription will work in general case as well.

The relationship  between  observables  $H_B$,  $H$  and $H^+$ in BFV,
refined algebraic and Dirac approaches is found.

Starting from BRST-BFV formula for  the  inner  product  and  obtained
relationship between  BRST-BFV  and  auxiliary  states,  we have found
expression \r{4} for  the  inner  product  of  auxiliary  states.  The
operator $\eta$  is  written  for general case of nontrivial structure
functions (eq.\r{16}).  Thus,  the  refined   algebraic   quantization
approach is generalized to the open-algebra case.

For a  simple  exactly  solvable example,  an explicit formula for the
inner product is obtained.

A wide class of such examples of systems with structure functions  can
be constructed as follows. Consider the Lie-algebra constrained system:
$\hat{\Lambda}_a = L_a - \frac{i}{2} f^c_{ac}$, $U^a_{bc} = f^a_{bc} =
const$ such that $L_a$ are linear in momenta,
$L_a = \alpha_{aj}(x) p_j + \beta_a(x)$.
The B-charge is
\bez
\Omega_0 =  C^a  L_a  - \frac{i}{2} f^a_{bc} \overline{\Pi}_a C^bC^c -
\frac{i}{2} f^a_{ba} C^b - i\pi_a\Pi^a.
\eez
Consider the unitary transformation being an exponent of the  operator
quadratic with respect to ghost variables,
\bez
U = \exp[\overline{\Pi}_a A^a_b(x) C^b - \frac{1}{2} A^a_a(x)]
\eez
It generates   a   linear  canonical  transformation  of  ghosts.  The
transformed B-charge $U^{-1}\Omega_0 U = \Omega$ is Hermitian
and nilpotent. It contains terms $\Omega^1$ and $\Omega^2$ only
and corresponds to the new system with classical constraints
\beq
\Lambda_{a'} = L_a (\exp A)^a_{a'}.
\l{27}
\eeq
with quantum corrections.
Generally, they form an algebra with nontrivial  structure  functions.
Since $\Omega^n=0$,  $n\ge3$,  while  the  constraints  are  linear in
momenta, the Cauchy problem analogous to \r{24} still  corresponds  to
the first-order  partial  differential  equation  and  can  be  solved
exactly, so that it is also possible  to  perform  an  integration  in
eq.\r{16} explicitly.

We see  that  the  system  with classical constraints \r{27} which was
mentioned in \c{Marolf3} can be exactly investigated by  the  approach
proposed in this paper.

The case   of  an  open  gauge  algebra  corresponding  to  nontrivial
coefficient functions $\Omega^n$,  $n\ge 3$ is much  more  complicated
for the  exact calculations.  However,  the integral formula \r{16} is
still valid,  so that one can use it for numerical calculations or for
application of  asymptotic  methods  such  as  perturbation  theory or
semiclassical approximation \c{Shvedov}.

The author is indebted to J.Klauder, Kh.S.Nirov, D.Marolf, V.A.Rubakov
and T.Strobl for helpful discussions.
This work was supported by the Russian Foundation for Basic  Research,
projects 01-01-06251 and 99-01-01198.

\end{document}